\documentstyle[prd,aps,preprint,tighten,epsfig]{revtex}

\begin{document}

\draft


\title{A Predictive Ansatz for Neutrino Mixing and Leptogenesis}
\author{\bf Zhi-zhong Xing}
\address{Institute of High Energy Physics, Chinese Academy of Sciences, \\
P.O. Box 918 (4), Beijing 100039, China \\
({\it Electronic address: xingzz@mail.ihep.ac.cn}) }
\maketitle

\begin{abstract}
We propose a non-SO(10) modification of the Buchm$\rm\ddot{u}$ller-Wyler
ansatz for neutrino mixing and leptogenesis, in which
charged lepton, Dirac and Majorana neutrino mass matrices have fewer
free parameters. Predictions of this new ansatz for three light
neutrino masses, three lepton flavor mixing angles,
the neutrinoless double-$\beta$ decay and the cosmological
matter-antimatter asymmetry are all in very good agreement with current
experimental and observational data.
\end{abstract}

\pacs{PACS number(s): 14.60.Pq, 13.10.+q, 25.30.Pt} 

\newpage

\section{Introduction}

The atmospheric and solar neutrino oscillations observed 
in the Super-Kamiokande \cite{SK} and SNO \cite{SNO} experiments have
provided rather convincing evidence that neutrinos are massive and lepton
flavors are mixed. To interpret the smallness of neutrino 
masses and the largeness of lepton flavor mixing angles, as indicated by 
current solar and atmospheric neutrino data, many theoretical models
and phenomenological ans$\rm\ddot{a}$tze have been 
proposed \cite{Review}. 
Among them, the one proposed by Buchm$\rm\ddot{u}$ller and Wyler \cite{BW}
is of particular interest, because (a) it is based on the simplest 
SO(10) lepton-quark mass 
relations \cite{SO10} and the seesaw mechanism \cite{Seesaw};
(b) it leads to quite specific predictions for the light neutrino masses, 
lepton flavor mixing angles, CP violation, and the neutrinoless
double-$\beta$ decay; and (c) it is able to
predict the cosmological matter-antimatter asymmetry via a very
attractive mechanism - leptogenesis \cite{FY}.

The present paper aims to propose a non-SO(10) 
modification of the Buchm$\rm\ddot{u}$ller-Wyler
ansatz, in order to account for current experimental 
data in a more accurate and more flexible way. 
To see why our effort makes sense, let us consider
the following relation obtained by Buchm$\rm\ddot{u}$ller and 
Wyler \cite{BW}:
\begin{equation}
\epsilon^2 \; \approx \; 
\frac{(1+ \tan^2\theta_{\rm atm})^3}{| \tan^2\theta_{\rm sun} -
\cot^2\theta_{\rm sun} |} \cdot \frac{\Delta m^2_{\rm sun}}
{\Delta m^2_{\rm atm}} \; ,
\end{equation}
where $\epsilon$ is a small expansion parameter of the Dirac
neutrino mass matrix; $(\Delta m^2_{\rm sun}, ~ \Delta m^2_{\rm atm})$
and $(\theta_{\rm sun}, ~ \theta_{\rm atm})$ are the mass-squared
differences and mixing angles of solar and atmospheric neutrino 
oscillations, respectively. From current Super-Kamiokande \cite{SK} 
and SNO \cite{SNO} data, one obtains 
\begin{eqnarray}
\Delta m^2_{\rm sun} & = & (3.3 - 17) \times 10^{-5} ~ {\rm eV}^2 \; ,
\nonumber \\
\tan^2 \theta_{\rm sun} & = & 0.30 - 0.58
\end{eqnarray}
at the $90\%$ confidence level \cite{Smirnov} 
\footnote{The best fit of current solar neutrino data 
in the large-angle Mikheyev-Smirnov-Wolfenstein (MSW) mechanism \cite{MSW}
yields $\Delta m^2_{\rm sun} = (5 - 7) \times 10^{-5} ~ {\rm eV}^2$
and $\tan^2 \theta_{\rm sun} = 0.35 - 0.45$ \cite{Smirnov2}.};
and 
\begin{eqnarray}
\Delta m^2_{\rm atm} & = & (1.6 - 3.9) \times 10^{-3} ~ {\rm eV}^2 \; ,
\nonumber \\
\sin^2 2\theta_{\rm atm} & > & 0.92
\end{eqnarray}
at the same confidence level \cite{Shiozawa}. Then it is 
straightforward to get $\epsilon > 0.1$ from Eqs. (1), (2) and (3).
This lower bound of $\epsilon$ is not consistent with  
$\epsilon \approx \sqrt{m_u/m_c} \approx 0.04 - 0.08$ 
obtained from the SO(10) relation between Dirac neutrino and up-type
quark mass matrices (i.e., $M_{\rm D} = M_{\rm u}$ \cite{BW}). Hence
one may wonder whether a modification of the 
Buchm$\rm\ddot{u}$ller-Wyler ansatz is possible, so as to avoid
any inconsistency with data or any fine-tuning of the parameter space.

It is obvious that the Buchm$\rm\ddot{u}$ller-Wyler ansatz will get much
flexibility to accommodate the present data on solar and atmospheric
neutrino oscillations, if the relevant SO(10) lepton-quark mass
relations are suspended. In this spirit, we propose a non-SO(10) 
modification of the Buchm$\rm\ddot{u}$ller-Wyler ansatz,
in which charged lepton, Dirac and Majorana neutrino mass matrices
have fewer free parameters. The predictability of such 
a new ansatz is expected to be more powerful. It is worthwhile to
emphasize that we are following a purely phenomenological approach. We 
hope that the good agreement of our results with current experimental 
and observational data may shed light on some appropriate ways of
model building, either within or beyond grand unified theories. 

The remainder of this paper is organized as follows. In section 2,
we specify the textures of charged lepton, Dirac and Majorana
neutrino mass matrices, from which one may see both similarities
and differences between the new ansatz and its original version.
Section 3 is devoted to explicit predictions of this new ansatz 
for light neutrino masses, lepton flavor mixing angles, CP violation, 
neutrino oscillations, and the neutrinoless double-$\beta$ decay. 
We calculate the lepton asymmetry in section 4,
and translate it into the baryon asymmetry via leptogenesis.
Finally, a brief summary is given in section 5.

\section{Lepton mass matrices}

A simple extension of the standard model is to include one
right-handed neutrino in each of three lepton families, while 
the Lagrangian of electroweak interactions keeps invariant under 
the $\rm SU(2)_{\rm L} \times U(1)_{\rm Y}$ gauge 
transformation \cite{FY}. After spontaneous symmetry breaking, the
lepton mass term can be written as
\begin{equation}
-{\cal L}_{\rm m} = 
\overline{(e ~ \mu ~ \tau )^{~}_{\rm L}} M_l 
\left (\matrix{ e \cr \mu \cr \tau \cr} \right )_{\rm R} + ~
\overline{(\nu_e ~ \nu_\mu ~ \nu_\tau )^{~}_{\rm L}} M_{\rm D} 
\left (\matrix{ \nu_e \cr \nu_\mu \cr \nu_\tau \cr} \right )_{\rm R} 
+ ~ \frac{1}{2} \overline{(\nu^{\rm c}_e ~ \nu^{\rm c}_\mu ~ 
\nu^{\rm c}_\tau )^{~}_{\rm L}} 
M_{\rm R} \left (\matrix{ \nu_e \cr \nu_\mu \cr \nu_\tau \cr} 
\right )_{\rm R} + ~ {\rm h.c.} \; ,
\end{equation}
where $\nu^{\rm c}_\alpha \equiv C \overline{\nu}^{\rm T}_\alpha$ with $C$
being the charge-conjugation operator (for $\alpha = e, \mu, \tau$); 
and $M_l$, $M_{\rm D}$ and
$M_{\rm R}$ stand respectively for the charged lepton, Dirac neutrino 
and Majorana neutrino mass matrices. We expect that the scale of $M_l$ 
and $M_{\rm D}$ is characterized by the gauge symmetry breaking scale
$v \approx 175$ GeV. The scale of $M_{\rm R}$ may be much higher than
$v$, because right-handed neutrinos are $\rm SU(2)_{\rm L}$ singlets
and their mass term is not subject to the electroweak symmetry
breaking. As a consequence, the $3\times 3$ light neutrino mass matrix 
$M_\nu$ arises from diagonalizing the $6\times 6$ neutrino mass matrix
\begin{equation}
{\bf M}_\nu ~ = ~ \left ( \matrix{
0	& M_{\rm D} \cr
M^{\rm T}_{\rm D}	& M_{\rm R} \cr} \right ) \; ,
\end{equation}
and takes the seesaw form \cite{Seesaw}
\begin{equation}
M_\nu ~ \approx~ - M_{\rm D} M^{-1}_{\rm R} M^{\rm T}_{\rm D} \; .
\end{equation}
Given specific textures of $M_{\rm D}$ and $M_{\rm R}$, one can
calculate the mass eigenvalues of $M_\nu$. The phenomenon of 
lepton flavor mixing at low energy scales stems from a non-trivial mismatch 
between diagonalizations of $M_l$ and $M_\nu$. In contrast, the 
lepton asymmetry at high energy scales depends on complex 
$M_{\rm D}$ and $M_{\rm R}$ \cite{FY}.

Now let us propose a non-SO(10) modification of the 
Buchm$\rm\ddot{u}$ller-Wyler ansatz \cite{BW} for neutrino mixing 
and leptogenesis. First of all, we assume $M_l$ and $M_{\rm D}$ 
to be symmetric matrices, just like $M_{\rm R}$. Second, we
assume that the (1,1), (1,3) and (3,1) elements of $M_l$,
$M_{\rm D}$ and $M_{\rm R}$ are all vanishing in a specific flavor basis, 
in analogy to a phenomenologically-favored 
texture of quark mass matrices $M_{\rm u}$ and $M_{\rm d}$ \cite{FX99}.
Note that we do not invoke any direct relationship between $(M_{\rm D}, M_l)$ 
and $(M_{\rm u}, M_{\rm d})$ (such as $M_{\rm D} = M_{\rm u}$ and
$M_l = M_{\rm d}$ in the SO(10) grand unified theory \cite{BW}).
Instead, we assume that the non-zero elements of $M_{\rm D}$ and $M_l$ 
can be expanded in terms of the Wolfenstein parameter 
$\lambda \approx 0.22$ \cite{Wolfenstein}. It is well known that the 
mass spectra of charged leptons and quarks are hierarchical \cite{PDG}:
\begin{eqnarray}
\frac{m_e}{m_\tau} & \sim & \lambda^6 \; , ~~~~~ 
\frac{m_\mu}{m_\tau} ~ \sim ~ \lambda^2 \; ;
\nonumber \\
\frac{m_u}{m_t} & \sim & \lambda^8 \; , ~~~~~
\frac{m_c}{m_t} ~ \sim ~ \lambda^4 \; ;
\nonumber \\
\frac{m_d}{m_b} & \sim & \lambda^4 \; , ~~~~~
\frac{m_s}{m_b} ~ \sim ~ \lambda^2 \; .
\end{eqnarray}
We conjecture that $M_{\rm D}$ might have a similar hierarchy as 
$M_{\rm d}$, but its dominant mass eigenvalue  
should be close to the electroweak symmetry breaking scale
$v \approx 175$ GeV. To be more explicit, we take
\begin{eqnarray}
M_{\rm D} & = & m^{~}_0 \left (\matrix{
0	& \hat{\lambda}^3	& 0 \cr
\hat{\lambda}^3	& x \hat{\lambda}^2	& \hat{\lambda}^2 \cr
0	& \hat{\lambda}^2 	& e^{i\zeta} \cr} \right )  ,
\nonumber \\
M_l & = & m^{~}_\tau \left (\matrix{
0	& \lambda^4	& 0 \cr
\lambda^4	& y \lambda^2	& \lambda^3 \cr
0	& \lambda^3 	& 1 \cr} \right )  ,
\end{eqnarray}
where $\hat{\lambda} \equiv \lambda e^{i\omega}$; $(x, y)$ are real and 
positive coefficients of ${\cal O}(1)$; and $m_0 \approx v$ holds. It is
easy to check that three mass eigenvalues of $M_{\rm D}$ have the hierarchy
$\lambda^4 : \lambda^2 : 1$, and those of $M_l$ have the hierarchy shown in
Eq. (7). The hierarchical structure of $M_l$ implies that its contribution
to lepton flavor mixing is very small and even negligible. Thus we expect that 
large lepton mixing angles observed in solar and atmospheric neutrino 
oscillations are essentially attributed to the light neutrino mass 
matrix $M_\nu$ in our ansatz.

Because the (1,1), (1,3) and (3,1) elements of both $M_{\rm R}$ and $M_{\rm D}$
have been assumed to be vanishing, $M_\nu$ must have the same texture zeros 
via the seesaw relation in Eq. (6) \cite{FXreview}.
To generate a sufficiently large mixing angle in the $\nu_\mu$-$\nu_\tau$ sector
to fit current Super-Kamiokande data on atmospheric neutrino 
oscillations \cite{SK}, the (2,2),
(2,3), (3,2) and (3,3) elements of $M_\nu$ should be comparable in magnitude.
This requirement is actually strong enough to constrain the texture 
of $M_{\rm R}$ in a quite unique way, as first observed by Buchm$\rm\ddot{u}$ller
and Wyler \cite{BW}. For our purpose, we obtain
\footnote{Note that $M_{\rm R}$ is given in terms of $\lambda$ rather than
$\hat{\lambda}$. If both $M_{\rm D}$ and $M_{\rm R}$ were expanded in terms of
$\hat{\lambda}$, the resultant texture of $M_\nu$ would be unable
to generate a large mixing angle in the $\nu_e$-$\nu_\mu$ sector.}
\begin{equation}
M_{\rm R} ~ = ~ M_0 \left ( \matrix{
0	& \lambda^5	& 0 \cr
\lambda^5	& z \lambda^4	& \lambda^4 \cr
0	& \lambda^4	& 1 \cr} \right ) ,
\end{equation}
where $z$ is a real and positive coefficient of ${\cal O}(1)$, 
and $M_0 \gg v$ holds. The texture of $M_\nu$ turns out to be
\begin{equation}
M_\nu ~ = ~ \frac{m^2_0}{M_0} \left ( \matrix{
0	& \hat{\lambda}		& 0 \cr
\hat{\lambda}	& z'	& 1 \cr
0	& 1	& e^{i2\varphi} \cr} \right ) ,
\end{equation}
where $z' \equiv 2x - ze^{i\omega}$ with $|z'| \sim {\cal O}(1)$,
and $2\varphi \equiv 2\zeta - 5\omega$. Note that
an overall phase factor $e^{i(5\omega-\pi)}$ has
been omitted from the right-hand side of Eq. (10), 
since it has no contribution to lepton flavor mixing
and CP violation at low energy scales. 

We remark that the (2,2), (2,3), (3,2) and (3,3) elements of $M_\nu$
in Eq. (10) are all of ${\cal O}(1)$, from which a large mixing
angle (around $\pi/4$) can be obtained for the $\nu_\mu$-$\nu_\tau$
sector. It is due to such a prerequisite that the hierarchy of $M_{\rm R}$
can almost uniquely be fixed through the seesaw relation between
$M_{\rm R}$ and $M_\nu$. In other words, we essentially require 
little information about the $\nu_e$-$\nu_\mu$ sector of $M_\nu$ to arrive
at Eq. (9)
\footnote{The $\nu_e$-$\nu_\mu$ sector of $M_\nu$ is generally
sensitive to the renormalization effects, in particular when the
corresponding mass eigenvalues ($m_1$ and $m_2$) are nearly 
degenerate \cite{RGE}.}.
As the $\nu_\mu$-$\nu_\tau$ sector of $M_\nu$ is relatively insensitive
to the renormalization effects from one scale to another \cite{RGE},
we expect that our phenomenological constraints on the texture of $M_{\rm R}$
at high energy scales make sense.
To generate a large mixing angle
in the $\nu_e$-$\nu_\mu$ sector to fit current Super-Kamiokande \cite{SK}
and SNO \cite{SNO} data on solar neutrino oscillations, the condition 
\begin{equation}
\left | z' e^{i2\varphi} - 1 \right | ~ \equiv ~ \delta ~ \sim ~ 
{\cal O}(\lambda)
\end{equation}
must be satisfied \cite{BW,Vissani}. Some instructive constraints on 
the parameter space of $x$, $z$, $\omega$ and
$\zeta$ can be drawn from Eq. (11), as one will see below. 

It is worthwhile at this point to comment on two major differences 
of the present ansatz from its original version \cite{BW}:

(a) We do not assume any SO(10) lepton-quark mass relations. Therefore
the pattern of $M_{\rm D}$ can be taken as Eq. (8) with a structural hierarchy 
weaker than before. This modification will lead to a ratio of 
$\Delta m^2_{\rm sun}$ to $\Delta m^2_{\rm atm}$ at the percent level
(i.e., $\epsilon^2$ in Eq. (1) is replaced by $\lambda^2$), consistent
very well with current experimental data. In this sense, we would say that
the phenomenological success of this new ansatz may compensate for the
theoretical cost for having discarded the simplest SO(10) mass relations.

(b) The number of free parameters in $M_{\rm D}$ and $M_{\rm R}$ is
reduced from ten \cite{BW} to five ($x$, $z$, $M_0$, $\omega$, and $\zeta$).
To do so, we have expanded $M_{\rm D}$ in terms of the complex parameter
$\hat{\lambda}$ and $M_{\rm R}$ in terms of the real
parameter $\lambda$. Such a treatment is plausible, since
the Majorana neutrino mass matrix $M_{\rm R}$ is {\it a priori} 
independent of the Dirac neutrino mass matrix $M_{\rm D}$. 
While $|\hat{\lambda}| =\lambda$ holds by definition, it actually
reflects the requirement of a large mixing angle in the 
$\nu_\mu$-$\nu_\tau$ sector of $M_\nu$, which imposes strict constraints
on $M_{\rm D}$ and $M_{\rm R}$ via the seesaw mechanism.
Because of the reduction of free parameters, the new ansatz is
expected to have more powerful predictability.

\section{Neutrino Mixing}

The symmetric neutrino mass matrix $M_\nu$ in Eq. (10) can be diagonalized by 
a $3\times 3$ unitary matrix $V$,
\begin{equation}
V^\dagger M_\nu V^* = \left ( \matrix{
m_1	& 0	& 0 \cr
0	& m_2	& 0 \cr
0	& 0	& m_3 \cr} \right ) ,
\end{equation}
where $m_1$, $m_2$ and $m_3$ are {\it physical} (real and positive)
masses of three light neutrinos. 
As pointed out above, the contribution of $M_l$ to lepton flavor mixing 
is expected to be very small and even negligible. Therefore the matrix $V$, which 
links the neutrino mass eigenstates ($\nu_1, \nu_2, \nu_3$) to the neutrino 
flavor eigenstates ($\nu_e, \nu_\mu, \nu_\tau$), can well describe the 
dominant effects of lepton flavor mixing at low energy scales. 
Current experimental data on solar, atmospheric and reactor 
neutrino oscillations \cite{SK,SNO,CHOOZ} strongly suggest that 
$|V_{e3}| \ll 1$, $|V_{e1}| \sim |V_{e2}|$ and $|V_{\mu 3}| \sim |V_{\tau 3}|$ 
hold. Then a parametrization of $V$ needs two big
mixing angles ($\theta_x$ and $\theta_y$) and one small mixing angle 
($\theta_z$) \cite{Xing02}, in addition to a few complex phases.
After some lengthy but straightfoward calculations, we obtain 
\footnote{Note again that an overall phase factor $e^{i(\pi - 5\omega)/2}$ has been 
omitted from the right-hand side of Eq. (13), in accord with Eq. (10).}
\begin{equation}
V \; \approx \; \left ( \matrix{
c_x e^{i(\alpha -\gamma+\pi/2)}	&
s_x e^{i(\alpha -\gamma)}	&
s_z e^{i\alpha} \cr
-s_x c_y e^{i(\beta +\gamma+\pi/2)}	&
c_x c_y e^{i(\beta +\gamma)}	&
s_y e^{i\beta} \cr
s_x s_y e^{i(\zeta +\gamma+\pi/2)}	&
-c_x s_y e^{i(\zeta +\gamma)}	&
c_y e^{i\zeta} \cr} \right ) ,
\end{equation}
in which $s_a \equiv \sin\theta_a$ and $c_a \equiv \cos\theta_a$ (for $a=x,y,z$),
$\alpha \equiv \omega + \zeta$, 
$\beta \equiv 5\omega - \zeta = \zeta - 2\varphi$, and
$2 \gamma \equiv \arg (z' e^{i2\varphi} -1)$. 
The explicit expressions of three mixing angles 
$(\theta_x, \theta_y, \theta_z)$ are 
\begin{eqnarray}
\theta_x & \approx & \frac{1}{2} \arctan \left ( 2\sqrt{2} \frac{\lambda}{\delta} 
\right ) \; ,
\nonumber \\
\theta_y & \approx & \frac{1}{2} \arctan \left ( \frac{2}{\delta} \right ) \; ,
\nonumber \\
\theta_z & \approx & \frac{1}{2} \arctan \left ( \frac{\lambda}{\sqrt{2}} 
\right ) \; .
\end{eqnarray}
In addition, three neutrino massses are given by
\begin{eqnarray}
m_1 & \approx & \left ( \frac{\lambda}{2\sqrt{2}} \tan \theta_x \right ) m_3 \; ,
\nonumber \\
m_2 & \approx & \left ( \frac{\lambda}{2\sqrt{2}} \cot \theta_x \right ) m_3 \; ,
\nonumber \\
m_3 & \approx & 2 \frac{m^2_0}{M_0} \; .
\end{eqnarray}
We can see that $\theta_z$ is as small as we have expected, and a normal 
neutrino mass hierarchy $m_1 : m_2 : m_3 \sim \lambda : \lambda : 1$ shows up.

It is worth mentioning that our instructive results for $(m_1, m_2, m_3)$
and $(\theta_x, \theta_y, \theta_z)$ will not get dramatic variations, if
arbitrary coefficients of ${\cal O}(1)$ are taken for those non-zero 
elements in $M_{\rm D}$ and $M_{\rm R}$. The reason is simply that the
hierarchical structures of $M_{\rm D}$ and $M_{\rm R}$ guarantee a stable
texture of $M_\nu$, from which the light neutrino masses and flavor mixing 
angles can straightforwardly be derived. For instance, a replacement 
$\lambda^4 \Longrightarrow A \lambda^4$ with $|A| \sim {\cal O}(1)$
for the (2,3) and (3,2) elements of $M_{\rm R}$ does not affect the pattern
of $M_\nu$ in the leading-order approximation \cite{BW}. Given a replacement
$\hat{\lambda}^2 \Longrightarrow B \hat{\lambda}^2$ with $|B| \sim {\cal O}(1)$
for the (2,3) and (3,2) elements of $M_{\rm D}$, the only variation of
$M_\nu$ is that its corresponding (2,3) and (3,2) elements change from 1 to $B$. 
In this case, we find 
\begin{eqnarray}
\theta_x & \approx & \frac{1}{2} \arctan \left [ 2 \sqrt{1 + |B|^2} ~ 
\frac{\lambda}{\delta} \right ] \; , 
\nonumber \\
\theta_y & \approx & \frac{1}{2} \arctan \left [ \frac{2|B|}{1-|B|^2 +\delta}
\right ] \; ,
\nonumber \\
\theta_z & \approx & \frac{1}{2} \arctan \left [ \frac{2|B|\lambda}
{\left (1+|B|^2 \right )^{3/2}} \right ] \; ;
\end{eqnarray}
and
\begin{eqnarray}
m_1 & \approx & \left [ \frac{\lambda}{\left ( 1+ |B|^2 \right )^{3/2}} \tan \theta_x 
\right ] m_3 \; ,
\nonumber \\
m_2 & \approx & \left [ \frac{\lambda}{\left (1 + |B|^2 \right )^{3/2}} \cot \theta_x 
\right ] m_3 \; ,
\nonumber \\
m_3 & \approx & \left (1 + |B|^2 \right ) \frac{m^2_0}{M_0} \; .
\end{eqnarray}
It is obvious that Eqs. (14) and (15) can be reproduced, respectively, 
from Eqs. (16) and (17) with the choice $|B| =1$. Therefore, small deviations of
$|B|$ from unity do not give rise to significant changes of the results 
obtained in Eqs. (14) and (15). Note that an arbitrary coefficient of ${\cal O}(1)$
for the (3,3) element of $M_{\rm D}$ or $M_{\rm R}$ can always be absorbed through
a redefinition of the mass scale $m_0$ or $M_0$. On the other hand, 
an arbitrary coefficient of ${\cal O}(1)$ for the (1,2) and (2,1) elements of 
$M_{\rm D}$ or $M_{\rm R}$ can also be absorbed via a redefinition of 
the perturbative parameter $\hat{\lambda}$ or $\lambda$. 

Some interesting implications of the simple results in Eqs. (14) and (15) 
are discussed in order.

(1) The hierarchy of three light neutrino masses allows us to determine the absolute
value of $m_3$ from the observed mass-squared difference of atmospheric
neutrino oscillations $\Delta m^2_{\rm atm} \equiv |m^2_3 - m^2_2| \approx m^2_3$.
Using the recent Super-Kamiokande data listed in Eq. (2),
we obtain
\begin{equation}
m_3 \; \approx \; \sqrt{\Delta m^2_{\rm atm}} \; \approx \; 
(4.0 - 6.2) \times 10^{-2} ~ {\rm eV} \; .
\end{equation}
Given $m_0 \approx v$ for the Dirac neutrino mass matrix $M_{\rm D}$, 
the mass scale of three heavy Majorana neutrinos turns out 
to be
\begin{equation}
M_0 \; \approx \; 2 \frac{v^2}{m_3} \approx \;
(4.9 - 7.6) \times 10^{14} ~ {\rm GeV} \; .
\end{equation}
We observe that this mass scale is not far away from the scale of grand unified
theories $\Lambda_{\rm GUT} \sim 10^{16}$ GeV.

(2) The small parameter $\delta$ defined in Eq. (11)
can well be constrained, if we take account of current experimental data on
the mass-squared difference of solar neutrino oscillations 
$\Delta m^2_{\rm sun} \equiv |m^2_2 - m^2_1|$ shown in Eq. (2).
As the ratio 
$R \equiv \Delta m^2_{\rm sun}/\Delta m^2_{\rm atm}$ is given by
\begin{equation}
R \approx \frac{\delta}{16} \sqrt{8 \lambda^2 + \delta^2} \; \approx 
(0.85 - 10.6) \times 10^{-2} \; ,
\end{equation} 
we obtain $\delta \approx 0.21 - 1.2$ for $\lambda \approx 0.22$. 
Note that $\delta > 0.5$ is apparently in conflict with our original
assumption $\delta \sim {\cal O}(\lambda)$ in Eq. (11).
Therefore the reasonable range of $\delta$ should be 
$\delta \approx 0.21 - 0.50$, which leads in turn to
$R \approx (0.85 - 2.5) \times 10^{-2}$. Subsequently we fix 
$\delta = \sqrt{2}\lambda \approx 0.31$ as a typical input.

(3) Using $\delta = \sqrt{2}\lambda$, we explicitly obtain
\begin{equation}
\theta_x \; \approx \; 31.7^\circ \; , ~~
\theta_y \; \approx \; 40.6^\circ \; , ~~
\theta_z \; \approx \; 4.4^\circ \; .
\end{equation}
To a good degree of accuracy, the mixing factors of solar, atmospheric and 
reactor neutrino oscillations are associated respectively with 
$\theta_x$, $\theta_y$ and $\theta_z$ \cite{FX01}. From 
Eq. (14) or Eq. (21), we get 
\begin{eqnarray}
\sin^2 2 \theta_{\rm sun} & \approx & \sin^2 2\theta_x \approx 
\frac{8 \lambda^2}{8\lambda^2 + \delta^2} \approx \; 0.8 \; ,
\nonumber \\
\sin^2 2 \theta_{\rm atm} & \approx & \sin^2 2\theta_y \approx 
\frac{4}{4 + \delta^2} \approx \; 0.98 \; ,
\nonumber \\
\sin^2 2 \theta_{\rm rea} & \approx & \sin^2 2\theta_z \approx 
\frac{\lambda^2}{2} \approx \; 0.024 \; .
\end{eqnarray}
Note that we have kept the $\delta$-induced correction to 
$\sin^2 2\theta_{\rm atm}$, in order to illustrate its small
departure from unity (maximal mixing).
The typical results in Eq. (22) are in good agreement with current 
Super-Kamiokande \cite{SK}, SNO \cite{SNO} and CHOOZ \cite{CHOOZ} data.

(4) Due to the mass hierarchy of three light neutrinos, our ansatz 
predicts a relatively small value for the effective mass term 
of the neutrinoless double-$\beta$ decay:
\begin{eqnarray}
\langle m\rangle_{ee} & \equiv & 
\sum_{k=1}^3 \left ( m_k V^2_{ek} \right )
\; \approx \; \frac{\lambda^2}{8} m_3 ~~
\nonumber \\
& \approx & (2.4 - 3.8) \times 10^{-4} ~ {\rm eV} \; ,
\end{eqnarray}
which seems hopeless to be detected in practice.
Indeed the present experimental upper bound is 
$\langle m\rangle_{ee} < 0.35$ eV at the 
$90\%$ confidence level \cite{Beta}.

(5) CP or T violation in normal neutrino oscillations is measured by 
a universal and rephasing-invariant parameter ${\cal J}$ \cite{Jarlskog}, 
which can be calculated as follows:
\begin{equation}
{\cal J} = \left | {\rm Im} (V_{e 2} V_{ \mu 3} 
V^*_{e 3} V^*_{\mu 2}) \right | \; \approx \; 
\frac{\lambda^2 \sin 2\gamma}{4 \sqrt{8\lambda^2 + \delta^2}} \; ,
\end{equation}
where
\begin{equation}
~~~~ \sin 2\gamma \;\; \approx \;\; \frac{2x}{\delta} \sin 2\varphi ~ - ~
\frac{z}{\delta} \sin (2\varphi + \omega) \; . ~~~~~ 
\end{equation}
Because of $x \sim z \gg \delta \sim \lambda$, a significant cancellation on
the right-hand side of Eq. (25) is naturally expected. There exists an
interesting parameter space, in which
\begin{equation}
x = \frac{1}{\sqrt{2}} \; , ~~ z = 1 + \sqrt{2} \lambda \; , ~~
\zeta = - \omega = \frac{\pi}{4} \; .
\end{equation}
Considering Eq. (11), one may easily check that $\delta = \sqrt{2} \lambda$ 
{\it does} hold  for the chosen values of $x$, $z$, $\omega$ and $\zeta$.
It is particularly amazing that $\sin 2\gamma = 1$ holds in this case. 
Therefore we obtain ${\cal J} \approx \lambda /(4\sqrt{10}) \approx 2\%$. 
CP violation at the percent level could be measured in the 
future at neutrino factories \cite{Factory}.

\section{Leptogenesis}

The symmetric neutrino mass matrix $M_{\rm R}$ in Eq. (9)
can be diagonalized by a $3\times 3$ unitary matrix $U$,
\begin{equation}
U^\dagger M_{\rm R} U^* = \left ( \matrix{
M_1	& 0	& 0 \cr
0	& M_2	& 0 \cr
0	& 0	& M_3 \cr} \right ) ,
\end{equation}
where $M_1$, $M_2$ and $M_3$ are {\it physical} (real and positive)
masses of three heavy Majorana neutrinos. In the leading-order
approximation, we obtain
\begin{eqnarray}
M_1 & \approx & \frac{\lambda^6}{z} M_0 \; ,
\nonumber \\
M_2 & \approx & z \lambda^4 M_0 \; ,
\nonumber \\
M_3 & \approx & M_0 \; ;
\end{eqnarray}
and
\begin{equation}
U \; \approx \; \left ( \matrix{
i	& \displaystyle\frac{\lambda}{z}	& 0 \cr\cr
-i \displaystyle\frac{\lambda}{z}	& 1	& \lambda^4 \cr\cr
i \displaystyle\frac{\lambda^5}{z}	& ~ - \lambda^4 ~ & 1 \cr\cr} 
\right ) \; . ~~
\end{equation}
One can see that the masses of three heavy Majorana neutrinos perform a clear
hierarchy. In view of Eq. (19), we arrive explicitly at
\begin{equation}
\left \{ M_1, M_2, M_3 \right \} \; \approx \; \left \{5.2 \times 10^{10}, ~ 
1.8 \times 10^{12}, ~ 6.0 \times 10^{14} \right \} ~ {\rm GeV} \; ,
\end{equation}
if $M_0 = 6.0 \times 10^{14} ~ {\rm GeV}$ and $z = 1 + \sqrt{2}\lambda$
are typically taken.

A lepton asymmetry may result from the interference 
between tree-level and one-loop amplitudes of the decay of the {\it lightest} 
heavy Majorana neutrino with mass $M_1$ \cite{FY}. 
This asymmetry can be expressed, in the physical basis where $M_{\rm R}$ 
is diagonal and the Dirac neutrino mass matrix takes the form 
$M_{\rm D} U^*$ instead of $M_{\rm D}$, as \cite{epsilon1}
\begin{equation}
\varepsilon_1 \; \approx \; - \frac{3}{16 \pi v^2} \cdot
\frac{M_1}{ [ U^{\rm T} M^\dagger_{\rm D} M_{\rm D} U^* ]_{11}} 
\sum^3_{j=2} \frac{{\rm Im} \left ( [ U^{\rm T} M^\dagger_{\rm D}
M_{\rm D} U^* ]_{1j} \right )^2}{M_j} \; ,
\end{equation}
where $v \approx 175$ GeV denotes the 
electroweak scale. In writing out Eq. (31), we have taken account of 
the strong mass hierarchy $M_1 \ll M_2 \ll M_3$.
With the help of Eqs. (8), (28) and (29), we get
\begin{equation}
\varepsilon_1 \; \approx \; - \frac{3 \lambda^6}{16 \pi} \cdot
\frac{x^2 z \sin 2\omega - 2x (1 + x^2) \sin \omega + \sin 2 (2\omega - \zeta)}
{z \left ( 1 + x^2 + z^2 - 2xz \cos\omega \right )} \; .
\end{equation}
Once the parameters $x$, $z$, $\omega$ and $\zeta$ are specified, one will be
able to predict the magnitude of $\varepsilon_1$ from Eq. (32).

For the purpose of illustration, we adopt the specific parameter space
given in Eq. (26) to evaluate the size of $\varepsilon_1$.
The result is
\begin{equation}
\varepsilon_1 \; \approx \; - \frac{\lambda^6}{4\pi} 
\left ( 1 ~ - ~ \frac{23\sqrt{2}}{12} \lambda ~ + ~ 
\frac{67}{18} \lambda^2 \right ) \; ;
\end{equation}
or numerically $\varepsilon_1 \approx - 5.2 \times 10^{-6}$.
To translate this lepton asymmetry into the baryon asymmetry of the universe \cite{FY}, 
one needs to calculate a suppression factor $\kappa$ induced by the 
lepton-number-violating wash-out processes \cite{Kolb}. 
Note that $\kappa$ depends closely on the following quantity:
\begin{eqnarray}
K_{\rm R} & \equiv & \frac{[U^{\rm T} M^\dagger_{\rm D} M_{\rm D} U^*]_{11}}
{8 \pi v^2} \cdot \frac{M_{\rm Pl}}{1.66 \sqrt{g^{~}_*} M_1} \; 
\nonumber \\
& \approx & \frac{3 - \sqrt{2} \lambda + 6\lambda^2}{16 \pi} \cdot
\frac{M_{\rm Pl}}{1.66 \sqrt{g^{~}_*} M_0} \; , ~
\end{eqnarray}
which characterizes the out-of-equilibrium decay rate of the {\it lightest}
heavy Majorana neutrino with mass $M_1$. 
In Eq. (34), $g^{~}_* \approx 100$ represents
the number of massless degree of freedom at the time of the decay,
and $M_{\rm Pl} \approx 1.22 \times 10^{19}$ GeV is the Planck mass scale.
Making use of the typical inputs taken above for Eq. (30), we arrive at
$K_{\rm R} \approx 73$. The suppression factor $\kappa$ can then be
calculated with the help of an approximate parametrization \cite{Kolb} 
obtained from integrating the Boltzmann equations
(for $10\leq K_{\rm R} \leq 10^6$):
\begin{equation}
\kappa \; \approx \; \frac{0.3}{K_{\rm R}}
\cdot \frac{1}{\left (\ln K_{\rm R} \right )^{0.6}}
\; \approx \; 1.7 \times 10^{-3} \; . ~
\end{equation}
Finally we get an instructive prediction for the asymmetry between baryon
($n^{~}_{\rm B}$) and anti-baryon ($n^{~}_{\rm\bar B}$) numbers of the
universe:
\begin{equation}
Y_{\rm B} \equiv  \frac{n^{~}_{\rm B} - n^{~}_{\rm\bar B}}{\bf s}
= \frac{c \kappa \varepsilon_1}{g^{~}_*}
\approx 4.7 \times 10^{-11} \; , ~
\end{equation}
where $\bf s$ denotes the entropy density, and $c = -8/15$
describes the fraction of $\varepsilon_1$ converted into $Y_{\rm B}$ via
sphaleron processes in the framework of three lepton-quark families and
two Higgs doublets \cite{Turner}. One can see that our result is 
consistent quite well with the observed baryon asymmetry, 
$Y_{\rm B} \approx (1 - 10) \times 10^{-11}$ \cite{Olive}.

Of course, one may go beyond the typical parameter space taken in Eq. (26) 
to make a delicate analysis of all measurables or observables, only if
the condition in Eq. (11) is satisfied. It is remarkable that we
can quantitatively interpret both the baryon asymmetry 
of the universe and the small mass-squared differences and large mixing 
factors of solar and atmospheric neutrino oscillations. 
In this sense, our ansatz is a {\it complete} phenomenological ansatz 
favored by current experimental and observational data, although 
it has not been incorporated into a convincing theoretical model.
A number of different ans$\rm\ddot{a}$tze on leptogenesis and
neutrino oscillations have recently been proposed \cite{Others,Review2}, but 
some of them turn to be ruled out by the present Super-Kamiokande and SNO data.

Future neutrino experiments will test the present ansatz and help to 
distinguish it from other viable models in the following four aspects:

(a) Our ansatz predicts the ratio of $\Delta m^2_{\rm sun}$ to
$\Delta m^2_{\rm atm}$ (i.e., $R$) to be around $1\%$. If more accurate solar
and atmospheric neutrino data yield $R < 0.5 \%$ or $R > 5\%$, our ansatz
will somehow become disfavored.

(b) The prediction of our ansatz for the smallest lepton mixing angle 
is quite certain: $\theta_z \approx \lambda/(2\sqrt{2})$ (or $4.4^\circ$).
This result may easily be examined in a variety of long-baseline neutrino 
oscillation experiments \cite{LBL}. Some other ans$\rm\ddot{a}$tze \cite{Others}
have given quite different predictions for $\theta_z$, either much larger
or much smaller than ours. 

(c) The magnitude of $\langle m\rangle_{ee}$ predicted by our ansatz 
(of order $10^{-4}$ eV) is too small to be measured in any proposed 
experiments for the neutrinoless double-$\beta$ decay \cite{Beta2}. 
If an unambiguous signal of the neutrinoless double-$\beta$ decay is observed 
in the near future, the present ansatz will definitely be ruled out. 

(d) In our ansatz, the rephasing-invariant parameter of CP violation 
(i.e., $\cal J$) is predicted to be at the percent level. This strength of
leptonic CP or T violation could be detected in the far future at neutrino 
factories \cite{Factory}. Therefore, another criterion to discriminate between 
our ansatz and other viable models is to see how large the CP-violating
effects can be in neutrino oscillations.

\section{Summary}

In summary, we have proposed a non-SO(10) modification of the 
Buchm$\rm\ddot{u}$ller-Wyler ansatz for neutrino mixing and leptogenesis.
Its consequences on the light neutrino masses, lepton flavor mixing
angles, the neutrinoless double-$\beta$ decay, and the baryon asymmetry
are all in good agreement with current experimental and observational
data. In particular, an indirect connection between the lepton asymmetry
at high energy scales and CP violation in neutrino oscillations has shown
up in such a specific ansatz. We expect that phenomenological
ans$\rm\ddot{a}$tze of this nature will get more stringent tests in the
era of long-baseline neutrino oscillation experiments. On the theoretical
side, a deeper understanding of the origin of fermion masses, flavor
mixing and CP violation becomes more desirable than before.

\acknowledgments{
The author would like to thank H. Fritzsch for 
warm hospitality in Universit$\rm\ddot{a}$t M$\rm\ddot{u}$nchen, where 
part of this work was done. He is also grateful to W. Buchm$\rm\ddot{u}$ller
for enlightening discussions in Bamberg, and to H. Fritzsch for 
helpful comments in Amsterdam.}

\end{document}